# The Dark Energy Survey Data Processing and Calibration System

Joseph J. Mohr[a,b], Robert Armstrong[c], Emmanuel Bertin[d], Gregory E. Daues[e], Shantanu Desai[a], Michelle Gower[e], Robert Gruendl[f], William Hanlon[g], Nikolay Kuropatkin[h], Huan Lin[h], John Marriner[h], Don Petravick[e], Ignacio Sevilla[i], Molly Swanson[j], Todd Tomashek[e], Douglas Tucker[h], and Brian Yanny[h] for the Dark Energy Survey Collaboration

[a]Ludwig-Maximilians University Department of Physics, Scheinerstr 1, Munich, Germany 81679; [b]Max Planck Institute for Extraterrestrial Physics, Giessenbachstrasse, Garching, Germany 85748; [c]University of Pennsylvania Department of Physics, 203 South 33rd St., Philadelphia, PA, USA 19104; [d]Institut d'Astrophysique, 98bis, bd Arago, Paris, France 75014; [e]National Center for Supercomputing Applications, 1205 West Clark St., Urbana, IL, USA 61801; [f]University of Illinois Department of Astronomy, 1002 West Green St, Urbana, IL, USA 61801; [g]University of Illinois Department of Physics, 1002 West Green St, Urbana, IL, USA 61801; [h]Fermi National Accelerator Laboratory, P. O. Box 500, Batavia, IL, USA 60510; [i]Centro de Investigaciones Energeticas Medioambiantales y Tocnologicas, Av. Complutense 40, Madrid, SP 28040; [j]Harvard Smithsonian Center for Astrophysics, 60 Garden St, Cambridge, MA USA 02138;

**ABSTRACT**

The Dark Energy Survey (DES) is a 5000 deg$^2$ grizY survey reaching characteristic photometric depths of 24$^{th}$ magnitude (10 sigma) and enabling accurate photometry and morphology of objects ten times fainter than in SDSS. Preparations for DES have included building a dedicated 3 deg$^2$ CCD camera (DECam), upgrading the existing CTIO Blanco 4m telescope and developing a new high performance computing (HPC) enabled data management system (DESDM).

The DESDM system will be used for processing, calibrating and serving the DES data. The total data volumes are high (~2PB), and so considerable effort has gone into designing an automated processing and quality control system. Special purpose image detrending and photometric calibration codes have been developed to meet the data quality requirements, while survey astrometric calibration, coaddition and cataloging rely on new extensions of the AstrOmatic codes which now include tools for PSF modeling, PSF homogenization, PSF corrected model fitting cataloging and joint model fitting across multiple input images.

The DESDM system has been deployed on dedicated development clusters and HPC systems in the US and Germany. An extensive program of testing with small rapid turn-around and larger campaign simulated datasets has been carried out. The system has also been tested on large real datasets, including Blanco Cosmology Survey data from the Mosaic2 camera. In Fall 2012 the DESDM system will be used for DECam commissioning, and, thereafter, the system will go into full science operations.

## 1. INTRODUCTION

The DES collaboration seeks to study the underlying cause of the cosmic acceleration through a combination of probes that include (1) galaxy cluster surveys[1,2,3], (2) cosmic shear[4], (3) galaxy clustering[5,6] and (4) SNe distances[7,8]. These studies will be carried out using the survey dataset, which will consist of a deep, multiband grizY survey over 5000 deg$^2$ together with ten dedicated deep SNe study fields, corresponding to 30 deg$^2$. The special purpose DES camera (DECam[9]) has been completed and shipped to CTIO where it is currently being installed. A variety of telescope improvements have been completed, including a new telescope control system and improved control of the primary mirror alignment. The data management (DESDM) development is still underway with the goal of improving various characteristics of the cataloging and addressing filesystem and database related bottlenecks that have been uncovered in large scale testing. Even with these remaining issues, the data products from small scale testing on real datasets (~10$^2$ deg$^2$) are of sufficient quality to be used for galaxy cluster science and photometric redshifts[10]. First light is scheduled for September 2012 and commissioning and science verification follow thereafter. If all goes as planned, the second half of the fall observing season will be dedicated to DES science operations.

In Section 2 we describe the characteristics of the DES dataset, which set the context for the DESDM system that is described in Section 3. In Section 4 we summarize the tests that have been carried out on DESDM using both simulated and real datasets, and in Section 5 we provide a summary and some further discussion.

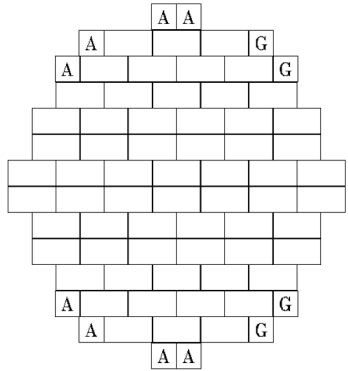

Figure 1: The DECam focal plane consists of 62 2048x4096 science CCDs distributed over a 2.2 degree diameter field of view. The DES imager has 3 deg$^2$ of active area corresponding to 520 million pixels, each subtending ~0.27 arcsec. Guider and auto-focus CCDs are marked with G and A.

## 2. CHARACTERISTICS OF THE DATASET

The characteristics of the dataset derive from the characteristics of the camera and the survey strategy. DECam[10] is a CCD camera consisting of 62 fully depleted, 250 micron thick 2048x4096 CCDs together with four 2048x2048 guider and eight 2048x2048 autofocus CCDs (see Figure 1). Each readout produces a ~1 GB image, and the readout occurs in under 17 seconds. DECam will be used with 5 filters grizY during the survey, and the thick, fully depleted CCDs lead to high quantum efficiency even in the z and Y bands.

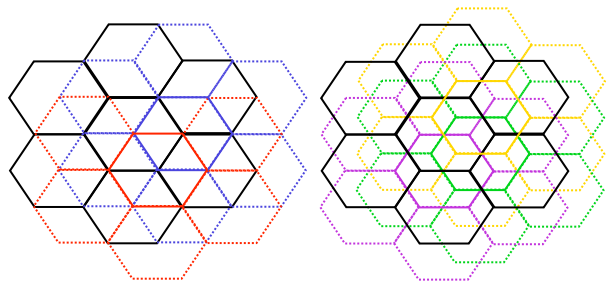

Figure 2: The DES tiling strategy where DECam pointings are represented as hexagons laid out in a honeycomb configuration on the sky. Rather than small dithers around each pointing as in the CFHT Legacy Survey, the DES survey strategy involves large offsets between overlapping layers to provide for greater uniformity in the dataset. (figure courtesy Jim Annis).

### 2.1 Survey Strategy

The survey strategy for the DES consists of repeated observations of 10 fields with small dithers to support the SNe science, but the large survey plan has two stages. First, an area of approximately 10% of the full survey solid angle of 5000 deg$^2$ will be imaged to full DES depth. This will allow us to immediately ascertain the data quality for the full survey and to move forward with exciting, near term science using these smaller areas. Second, the remaining areas of the survey will be imaged over successive seasons, building up approximately 2 layers of imaging in each band during each season. The DES tiling strategy involves creating a full layer of imaging in a single band covering the full 5000 deg$^2$ area. Within each layer of imaging there are only minimal overlaps between neighboring fields. Subsequent layers of images are then offset from the last by approximately half a field of view (see Figure 2). These shifts between layers have been optimized to avoid overlapping chip gaps between layers. This approach leads to every source in the survey having been observed using approximately ten different parts of the focal plane, presumably allowing for more uniform data quality across the survey. Common stellar objects in overlapping images will be used to bring all observations within a band to a common zeropoint during image coaddition.

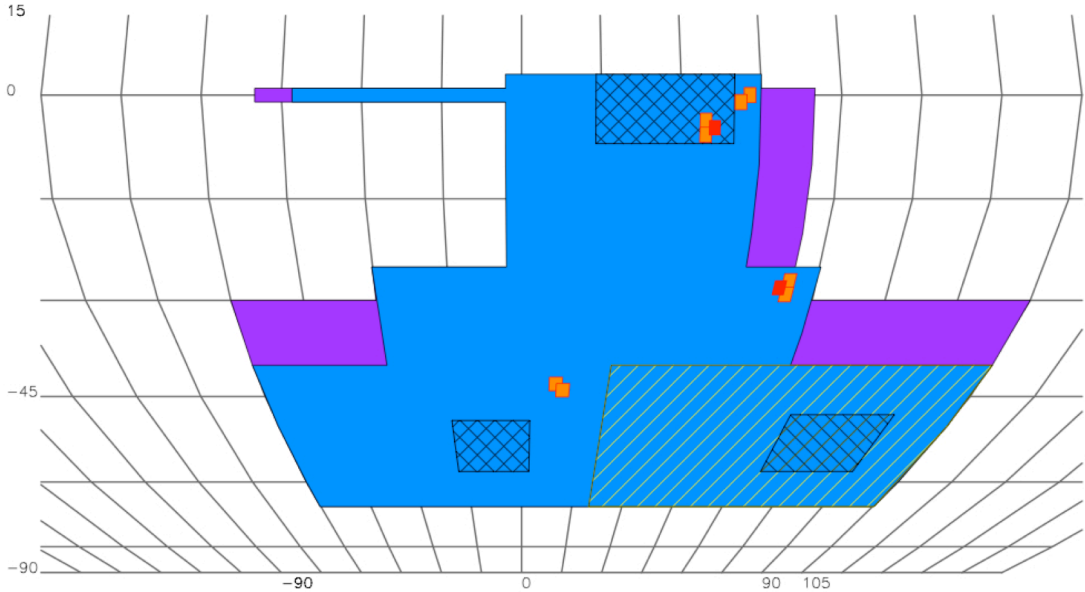

Figure 3: The current layout of the DES main survey is shown in blue above. SNe deep fields are marked in orange and red, and purple area is that which has been redistributed into the blue connecting region between the SPT fields (southern) and the equatorial strip. Hatched regions correspond to target full-depth early survey fields that are associated with the SPT-pol survey and equatorial regions important for eBOSS. The cyan box in the middle outlines the deep NIR VIKING survey that will be covered in full by DES (figure courtesy Jim Annis).

| *Band* | $t_{exp}$ [s] | $M_{lim}$ PSF [5$\sigma$] | $M_{lim}$ Galaxy [10$\sigma$] | FWHM ["] |
|---|---|---|---|---|
| *g* | 800 | 26.5 | 25.2 | 0.83 |
| *r* | 800 | 26.0 | 24.8 | 0.79 |
| *i* | 1000 | 25.3 | 24.0 | 0.79 |
| *z* | 1000 | 24.7 | 23.4 | 0.78 |
| *Y* | 500 | 23.0 | 21.7 | 0.77 |

Table 1: DES wide survey exposures together with estimated typical photometric depths and image quality.

The current survey strategy focuses in the south galactic cap region which is also the location of the 2500 deg$^2$ South Pole Telescope survey[11]. Each observing season runs from September through February for 125 nights, providing a total of 525 guaranteed nights spread over 5 seasons. The long term goal of DES is to continue the survey after the initial 5 years to complete the dataset over the full southern extragalactic cap region, but this goal will be reassessed as the initial 5 years survey comes to an end, taking the development timescale for LSST into account. Figure 3 shows the current survey plan, where the blue region corresponds to the planned wide survey and the SNe fields are marked in orange (shallow) and red (deep). The purple region marks area that was originally targeted for survey but that was then redistributed to enlarge the coverage of the cap region near the celestial equator. The cross hatched boxes correspond to about 10% of the wide survey region and are areas that will be observed to full depth during the first or second seasons. These full depth regions will give us enough data to test and tune our algorithms in preparation for the processing and analysis of the full survey. In addition, we expect these regions to enable exiting near term science. The hatched region on the bottom right corresponds to the portion of the SPT survey that we expect to observe to a standard single season depth of two exposures in each band.

## 2.2 Photometric Depths

Photometric depths for the DES survey have been estimated using the planned observing times in each band, the characteristics of the site, the measured QE of the CCDs and the throughput curves of the filters. The assumed PSF is adopted by using the expected performance of the new corrector together with the delivered site and telescope seeing. The median seeing is better than that obtained in the BCS survey using the Mosaic2 camera[10], but the hope is that the performance of the new DECam corrector will be significantly better than that of the old prime focus corrector. Table 1 presents the estimated 5 sigma PSF depths and 10 sigma galaxy depths together with the median seeing from the simulations. The galaxy depths are calculated within apertures that are 1.6 times the PSF FWHM.

## 2.3 Data Volumes and Data Model

A clear night of DECam observing is expected to produce about 300 science exposures together with approximately 60 bias and dome flat calibrations. Thus, the maximal dataset from the survey would be ~200K exposures, of which about 150K would be science exposures. The raw DES dataset is then <200TB. Out of these data we produce reduced images, coadd images and associated catalogs and quality assurance figures. All our image data products are stored in multi-extension FITS files, with each image associated with a pixel level weight map and a pixel level bad pixel map. In the case of the coadd images, we store two different weight maps, where the corrections for the correlated noise are applied for two different scales to enable accurate object detection and accurate measurement uncertainties. A reduced image is the size of a single CCD and corresponds to an astrometrically calibrated, detrended image of the sky; this image has not been remapped and so the noise is independent in each pixel. There are 62 of these for each survey science exposure, and the full dataset of compressed, reduced images should be about 130TB. We employ lossy Rice tile compression, assigning four bits to the noise, and we find that this provides about a factor of 5 compression for our dataset. The coadds are built in each band along with a multi-band detection image; each has an image, two weight maps and a bad pixel map, and so they require around 15TB for the wide survey (assuming 5x lossy compression as above). Overall, the DES source, single epoch and coadd datasets after a single processing would correspond to < 0.5 PB. Given that our baseline plan is to process the full survey after each season we plan for approximately four times more data in the end.

## 3. PROCESSING AND CALIBRATION

DESDM is an automated processing system that is built around a metadata repository. More details of the overall structure of the system can be found elsewhere[12,13]; here we provide only a short overview and then detail some elements of the processing and calibration that are of particular interest. The bulk of the processing can be carried out in a data parallel fashion: (1) each of the 62 CCD images within an exposure can be processed through most steps of detrending and calibration independently of the rest of the images and (2) coaddition and cataloging can proceed independently tile by tile on the sky. Our system is designed to capitalize on this data parallel capability by parceling the data out into independent jobs running in parallel. Our system is designed for use on either a dedicated cluster or on a high performance computing platform such as those in the US XSEDE program or those here in Europe at facilities such as the Leibniz Rechenzentrum (LRZ) here in Munich. Moreover, it is designed to support processing on multiple platforms, making it possible to quickly switch from one platform to another as maintenance requires or even, in principle, to run simultaneously on many platforms. While it is simplest to operate on a dedicated cluster, we have designed in the ability to run on public supercomputers so that during times of reprocessing it will be possible to push the data through our system at more than a order of magnitude higher rates than during our standard daily processing that takes place while the survey is in operation.

### 3.1 Detrending of Single Epoch Images

The raw data from the camera are split into individual CCD images and a crosstalk correction is applied. The crosstalk coefficients are extracted from sets of exposures by examining (in each independent combination of amplifiers) the relationship between the pixel flux in the primary amplifier to the pixel flux in the cross-talk affected amplifier. We examine pixel flux in the primary amplifier only above some threshold value, and so typically one is examining whether a 10,000 ADU pixel in the primary amplifier is reflected as a ~100 ADU source pixel in the crosstalk affected amplifier. In pairs where there is no evidence of crosstalk a coefficient of zero is returned. Typical measured crosstalk values in the DECam devices are at the 1% level, and the crosstalk is restricted to neighboring amplifiers that share common electronics.

Thereafter, the nightly calibration data are used to produce a bias correction image and a dome flat correction image in each band. In this process, a bad pixel map stored in the calibration portion of the archive is used to mask out hot columns, glowing edges and other pixel defects. In addition, each image is overscan corrected using a simple median value for each readout row. In the production of the flat field images, a pupil ghost correction is applied to the flats that accounts for scattered light resulting from multiple reflections within the five element DECam corrector. This additive pupil ghost is measured using star flats determined from calibrated data processed without the pupil ghost correction. We use the star flats to determine a functional form that represents the correction needed as a function of position to bring all stars throughout a detector to a common photometric zeropoint. This correction scales as a function of the sky brightness (i.e. a measure of the total light coming through the corrector in each band). No pupil ghost has been measured for DECam to date, so we are using a theoretical expectation extracted for the corrector optics and representing the shape as a broad Gaussian centered in the field. This correction is stored as 62 CCD correction images, and so our pupil ghost will work well even if the shape of the scattered light is quite complex.

The science exposures are then overscan corrected, bias corrected, flat field and pupil ghost corrected. A bad pixel image stored as a calibration product in the archive is used to build an initial bad pixel map, and additional pixels are marked as saturated as appropriate. By default we interpolate over single bad columns, enhancing the noise stored in the weight map by a factor of four to reflect the loss of information and marking these pixels as interpolated in the bad pixel mask. We typically carry out a round of cosmic ray masking and bright star masking at this stage as well.

Because we wish to avoid correlated noise in our reduced images, we do not remap them to portions of a tangent plane. Thus, we also have to correct the source flux for the pixel solid angle variation across each chip. To do this we use an archive calibration product that stores a pixel area scale factor image for each CCD. Because we wish to maintain the flat sky brightness that results from our dome flattening, we apply the pixel area correction to the modal sky subtracted image. In this way, the correction is only applied to the source flux, and after the correction the modal sky value for the image is added back. In principle, at this stage our reduced images have both flat sky *and* a non-varying photometric zeropoint.

Reduced images from within a common band and CCD are then combined with object pixel rejection to determine an illumination correction and a fringe correction (in z and Y bands). For our simulated datasets the illumination corrections and fringe corrections just contain noise fluctuations around one and zero, respectively. In real datasets the fringe correction, whose shape is determined as the common pattern across the full night of corrected images in a band, is a subtractive correction whose amplitude scales with the modal value of the sky. This works quite well for real datasets (namely from the Mosaic2 and the VLT FORS2 cameras). If we have determined the pupil ghost correctly from the data, then there should be no further need to flatten using an illumination correction. Testing with CFHT Legacy Survey data on this issue is still underway.

## 3.2 Astrometric Calibration

We use the AstrOmatic code SCAMP[14,15] along with the 2MASS catalog for astrometric calibration. SCAMP has been used extensively in the CFHT Legacy Survey, and its excellent performance has long been well established. For DESDM we have made some improvements. For one, within our processing we prefer to calibrate each exposure (collection of 62 reduced images from single readout) independently of the rest. For the DECam focal plane we find a good description of the distorted tangent plane is possible when we assign each CCD a $3^{rd}$ order polynomial distortion function around the tangent plane for both Right Ascension (RA) and Declination (DEC) plus another second order function over the entire focal plane to describe differential refraction. However, the accuracy of the solutions suffers in cases whether there are fewer sources on a chip due to a bright star. SCAMP was designed to be used on a large ensemble of overlapping exposures simultaneously, so that the common form of the high order CCD level distortions could be well determined using a large number of constraints. What we have done is introduce a facility that allows the tangent plane distortions common to the CCDs in all exposures within a band to be taken as input during the astrometric refinement. Then only the second order distortion parameters over the full focal plane are required from each exposure, and this can be handled robustly even when calibrating a single exposure. In a second improvement, the differential diffraction across the field is modeled more accurately so that SCAMP now operates reliably for very large field of view cameras like DECam even at high air mass.

### 3.3 Photometric Calibration

DESDM is designed to enable us to calibrate photometrically in a traditional way that uses nightly photometric solutions extracted from standard star observations taken under photometric conditions. Single epoch observations from fields where there are overlapping calibration catalogs contribute constraints to the nightly photometric solutions in each band. We solve for photometric solutions that include a common extinction coefficient across all CCDs but different zeropoints for each CCD. Color terms can be extracted across either the whole camera or in each CCD individually. These photometric solutions are ingested into the DESDM database and can then be used to update the zeropoints for all single epoch images from that night (as well as the photometry in catalogs that have already been extracted for those images).

This approach works well if there is a precise way to monitor the photometric quality of the night. To this end DES has built an IR cloud monitor that will be operated at CTIO. In tests on real data where observers have called nights photometric, we have found that erroneous judgments about whether a night is photometric have contributed considerable noise to the catalog photometry. This has led us to develop a separate track for photometric calibration of real data that seems to offer robustness against errors in the nightly photometric solution. In this approach we use common stellar objects in overlapping single epoch images to first bring all images within a band to a common zeropoint. We typically carry this out on each coadd tile independently, but in fact we have developed the code to handle the entire survey dataset simultaneously. To solve for the absolute zeropoint we use all the bands together and leverage the constancy of the stellar locus across the extragalactic sky together with the 2MASS NIR calibration, which is accurate at the 2% level. We have had good experience with this stellar locus calibration on tests carried out with real data[10].

The requirements on our zeropoint accuracy in DES are 2%, and so we cannot expect to meet this using only the stellar locus and calibration to 2MASS over small patches of the sky. Indeed, we intend to pursue a traditional calibration using all available photometric data and the large ensemble of stellar objects common to more than one single epoch survey image. The stellar locus can serve as an independent cross check of galaxy colors, and perhaps it will also be used for more.

### 3.4 PSF Modeling and Model Fitting Photometry

As we process data, we model the PSF and its variation across each single epoch image using the AstrOmatic code PSFEx developed by E. Bertin. This code takes NxN pixel stellar image cutouts from across an image and models the PSF variation as a polynomial expansion of CCD x and y position. For real and simulated data we find that allowing for second order variation is adequate. The PSF itself is then expressed as a sum of NxN pixel components $PSF^a(i,j)$ where each is weighted by the appropriate factor in the polynomial expansion:

$$PSF(x,y) = PSF^0(i,j) + x \cdot PSF^1(i,j) + y \cdot PSF^2(i,j) + xy \cdot PSF^3(i,j) + x^2 \cdot PSF^4(i,j) + y^2 \cdot PSF^5(i,j)$$

where (x,y) is the position on the CCD, (i,j) is the pixel location within each NxN PSF component and different components $PSF^a$ are used for each term in the polynomial expansion.

The AstrOmatic cataloging code SExtractor takes these PSFEx models as input and uses them to carry out PSF corrected model fitting photometry to all sources in an image. Within DESDM we extract PSF magnitudes as well as either elliptical Sersic for bulge plus disk models for all objects. As expected, the PSF magnitudes produce biased estimates of galaxy photometry, but the Sersic and bulge+disk galaxy models produce unbiased magnitudes of stars. In addition, with an accurate model of the PSF it is possible to extract much more accurate classification information using the new SExtractor classifier *SPREAD_MODEL*. In tests on real data we have found this classifier to produce smoothly varying distributions of stars on the sky to greater depth and over a broader range in image quality than the traditional *CLASS_STAR*[10].

Using PSFEx for PSF modeling and SExtractor to produce PSF corrected model fitting photometry then insulates the catalogs from biases associated with seeing variations from exposure to exposure and within each exposure as a function of position on the focal plane. These powerful codes provide a general purpose solution to this perennial problem in observational astronomy, and the solution is fast and robust enough to be carried out on all survey images. SExtractor running in PSF model fitting model typically catalogs ~10 objects/second, requiring a few minutes to catalog a typical

DECam reduced image. All reduced images can be processed independently of one another, and so with a modest cluster there are no timing concerns with this software.

### 3.5 Coaddition with PSF Homogenization

A typical challenge in real datasets is that the image quality can vary among the single epoch images that one combines into a coadd image. This can be minimized by obtaining all the images one right after another under similar conditions, but typically this requirement is difficult to meet for a large survey and for ground based observatories this strategy then still leads to inhomogeneity in the PSF from location to location on the sky. For the DES observing strategy where each object is observed in ~10 different locations on the focal plane and visits to the field are spread uniformly over a five year period, we face a situation where if we simply coadd the images we will end up with discontinuities in the PSF variation as a function of position within the coadd image. While we could model these discontinuous PSF variations using a high order polynomial in PSFEX, we could not model them accurately at the discontinuities. An approach like this would inevitably lead to systematics in the PSF corrected photometry and classification of galaxies and stars as a function of position within each coadd image.

To resolve this issue we have built tools that allow us to homogenize the PSF within the coadd images. With a homogenized (i.e. constant) PSF in the coadd, the PSFEx model for the PSF becomes trivial and accurate, and these photometric and classification biases go away. It works in the following way. First, we use the seeing across the ensemble of single epoch images to choose a target PSF for the homogenized coadd. We typically choose the median seeing within each band, in which case the PSF is being degraded in half the images and improved in the other half. Our system is set up to allow the user to choose any target PSF, including the worst seeing in the ensemble or, for example, the seeing which corresponds to 90% of the single epoch images having better seeing. Note that we homogenize PSFs within each band separately, because with the PSF corrected model fitting photometry available in SExtractor there is no benefit to bringing all bands to a common PSF.

Second, we run PSFEx together with a target PSF (i.e. Moffat with particular FWHM) in a different mode on each single epoch image to produce a position dependent smoothing kernel that, when used to convolve the image, results in the PSF being the target PSF at all locations in the image. We run this on the full ensemble of images to build the smoothing kernels and run a code called PSFnormalize that applies these kernels to the full ensemble, allowing them to then be coadded using the AstrOmatic code SWarp. This produces a coadd image with a homogeneous PSF that is ready for cataloging.

We have carried out extensive tests on simulated and real data of the performance of this tool. The noise correlations introduced by the PSF homogenization are more severe than those introduced by the remapping carried out in the SWarp coaddition. The noise impact is differential, amplifying the sky noise in images whose PSF has been improved and reducing it in images whose PSF has been degraded. This noise amplification/suppression is a function of scale and position within each image, with the noise being unaffected on scales several times larger than the PSF. We account for these changes in the noise within the weight maps that come with each image, and then during the coaddition process if one is using inverse variance weighted combination, these weights are used in an optimal fashion to produce the final coadd; that is, images where the noise has been amplified (because they started out with seeing worse than the target seeing) will be deweighted in the coaddition. To maintain accurate photometric measurement uncertainties and to maintain good object detection, we currently calculate two weight maps for the homogenized coadds. One includes the corrections for correlated noise on the scale of a pixel, and the other on the scale of the PSF. Our experiments with this approach are ongoing, but results are encouraging.

Because of the correlated noise issues associated with PSF homogenization, we also will produce direct coadds over the full survey. We have no plans to catalog these direct coadds, because of the PSF induced biases described above, but we expect these images to be quite useful to produce pretty pictures and in searching for low surface brightness objects such as strong lensing arcs in the cores of clusters.

### 3.6 Two Strategies for Deep Catalogs

To catalog the coadds we use SExtractor in dual image mode. The detection image is an optimal combination of the images from each DES band *grizY*. We will also carry out combined photometry with our partner survey VHS-DES, and we expect in that case to carry out cataloging using a detection image built from *grizYJHK*. We build PSF models for each band, and the resulting catalogs therefore include PSF corrected photometry for a common set of objects across all bands. Higher signal to noise galaxy colors can be extracted using small apertures that have been aperture corrected using the comparison to stellar PSF magnitudes or using a newly developed *DET_MODEL* magnitude within SExtractor[16]. This photometric measurement extracts a single PSF corrected model from the detection image and uses it to determine the photometry in each band using the appropriate PSF correction in each. Because the galaxy shape is not a free parameter in this case the statistical noise of the measurement is reduced. Initial tests show that, as expected, this approach produces higher signal to noise photometry for faint objects, but we are still working to characterize the biases.

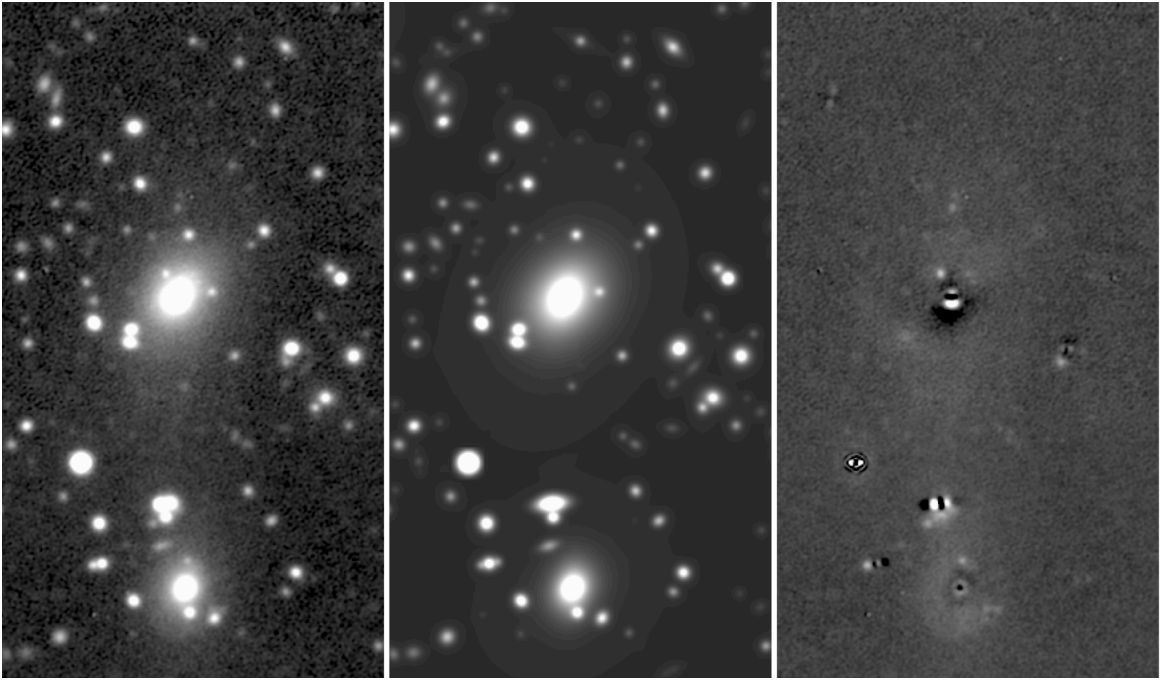

Figure 4: An image (left) of a cluster core from the BCS survey[10] with an associated model image (middle) and residual image (right). This example shows the results of four iterations of SExtractor model fitting photometry where additional components were added as needed. While there is still need for further work, the performance of model fitting for deblending in this complex field is far superior to single pass, non-model fitting codes.

While the strategy described above is the strategy we will take into operations this fall, we are pursuing a separate development and testing stream of SExtractor that will lead to the ability to carry out photometry over an ensemble of single epoch images simultaneously. In this approach the coadd images would be used to provide a deep detection image (if used at all), while the PSF corrected model fitting photometry will be carried out across multiple images at once. Given our approach of avoiding all remapping on the single epoch images, this approach to cataloging would allow for the optimal use the PSF information and pixel by pixel (uncorrelated) noise in all the images. When this code has been adequately tested we expect to transition to its use within DES operations.

Finally, it may be helpful to point out one additional benefit of model fitting photometry. It allows for the deblending of sources, effectively assigning each photon in regions where light distributions from multiple sources are blended to the associated highest probability source. Initial tests with SExtractor PSF corrected model fitting photometry show that this approach with multiple iterations of model fitting shows considerable promise in helping to resolve the perennial deblending challenges in cataloging codes.

## 4. TESTING AND VALIDATION

DESDM has benefited from a vigorous testing program on quite sophisticated simulated data. In addition, we have been using the system to process real data throughout the development period. Below we describe both elements of our testing program.

## 4.1 Tests with Simulated Data

Through 2009 we carried out yearly data challenges where we processed, calibrated and served 200 deg$^2$ of full depth DES survey data at NCSA. In terms of solid angle of imaging this is equivalent to about 20% of the SDSS dataset. The simulated DECam images that serve as input for our data challenges were produced by Huan Lin's team at Fermilab. These data challenges helped us to identify bottlenecks and to get feedback from the collaboration on our data access tools. Since 2009 only a single new data challenge DC6 has been carried out, but the data products were subjected to higher level scientific scrutiny through a process led by the DES science working groups.

At the point of the last regular data challenge DC5 in 2009, two bottlenecks to the processing had been identified. The first was the database, which was unable to support more than about 50 jobs running simultaneously. The second was a large parallel filesystem issue (minimum blocksize) that led the harvesting of metadata from new FITS data products from a processing step to take as much as ten times longer than on our servers. Since that time we have upgraded the database hardware at NCSA, and this has removed our database bottleneck for the time being. We don't yet have a solid estimate of the maximum number of jobs that we can run simultaneously without the database performance becoming an issue (Note that interactions with the database are required only during job setup and completion, but nevertheless, the load on the database was substantial even with 50 jobs running simultaneously). For the filesystem problem there is no easy solution given that large parallel filesystems are typically optimized for large files. This means that even if one is retrieving 10KB of metadata from a header, one must pay the price of loading the whole block. In the face of a multi-extension FITS file with $10^2$ extensions, this read process is unnecessarily slow. A different approach to collecting metadata and changes to the data model may well be required.

From the data quality perspective we have been meeting our core astrometric and photometric zeropoint accuracy requirements for several years now. However, DC6 uncovered a range of remaining data quality issues. Most of these fall into the category of cataloging issues, with the primary concern being large photometric errors on very bright (i.e. large) galaxies and larger than expected photometric scatter in faint galaxies. For the large galaxy problem we are looking at adjustments to our background grid during both coaddition and cataloging. This is not a serious challenge, although tuning these background parameters to improve cataloging of large galaxies may indeed have unintended consequences on our faint galaxy photometry in cases where the sky is not flat. For the faint galaxy photometry the real challenge is to extract the highest possible signal to noise color. As mentioned above in Section 3.6, we are exploring the new *DET_MODEL* estimator, and of course we can use small apertures to help with the galaxy colors on the faint end.

Our basic code development and testing program is supported by the processing of much smaller test datasets. We have developed the concept of a Gold Standard Night (GSN) that is about 20% the size of a single DES survey night. The simulated images are arranged so that we can produce an approximately full depth coadd along with several surrounding tiles that don't reach full depth. This dataset can be processed in less than a day on our dedicated servers, so it is possible to implement a code fix or parameter change and return statistics from a GSN test by the next day. The rapid turnaround of GSN has been a central element in the development of our science codes.

## 4.2 Tests with Real Data

To date the primary test with real data has been in processing and calibrating the Blanco Cosmology Survey (BCS)[10], which is an 80 deg$^2$ *griz* survey carried out with the Blanco 4m telescope at CTIO. The details of this testing program appear in the recent BCS paper. In summary, we photometrically calibrated our data using the stellar locus and saw excellent performance on single galaxy photometric redshifts using ANNz. In these photometric redshifts we used model magnitudes, small aperture galaxy colors, and the *SPREAD_MODEL* classifier which helps differentiate compact from diffuse galaxies to deliver characteristic scatter of <5% in δz/(1+z) out to z=1[10]. In addition, we have used the DESDM system to process a range of single cluster followup imaging data we have acquired within the South Pole Telescope (SPT) collaboration. This is quite an interesting test, because the image quality can be extremely poor. These data have been used for either photometric redshift cross-checks or as the primary source for photometric redshifts in all the SPT optical work, and we have achieved cluster photometric redshift accuracy of δz/(1+z)=0.017 out to z=1.32. Despite these scientific successes, we have not achieved the required DES photometric quality with the tests on real data.

At present we are pursuing a large scale test on the CFHT Legacy Survey COSMOS field. This dataset is quite well understood, and it has been used to extract excellent photometry accurate at ~1% level[17]. Moreover, there is a large spectroscopic training set available that we can use as a direct test of the accuracy of our photometry. In a first phase we will apply our PSF homogenization tools and carry our PSF corrected model fitting photometry to see how our photometry performs. In a second phase we will start from the raw data and attempt to extract the pupil ghost correction and apply it to see if we can match the high accuracy photometry currently being produced by Terapix. In addition to these tests on well understood CFHT Legacy Survey data we are also processing Pan-STARRS1 data in the COSMOS area. While the PS1 data are not nearly as well understood as the CFHT data, we expect to learn things from applying our tools to the 7 deg$^2$ field PS1 data.

## 5. SUMMARY AND PROSPECTS

In summary, we have developed a data management system that is designed for fast reprocessing of large datasets like DES using publicly available supercomputing facilities. We plan to operate our system for DES on a small dedicated cluster to support day to day processing during survey operations. We have developed DES specific tools for detrending of images, and we have relied on extensions of well tested AstrOmatic codes to provide astrometric calibration, image coaddition, PSF modeling and PSF corrected model fitting photometry. We have developed tools for the homogenization of images being combined into deep coadd images so that it is possible to carry out unbiased model fitting photometry over these coadds. The use of PSF corrected model fitting photometry provides advantages in dealing with inhomogeneous datasets and in the deblending of astrophysical sources in complex fields.

These codes have been extensively tested using simulated DECam images together with real data from the Mosaic2 camera, and additional real world testing is underway on CFHT Megacam data and Pan-STARRS1 data. In addition, we routinely use our system to process images from VLT-FORS2, ESO 2.2m-WFI, NTT-EFOSC, SOAR-SOI, Magellan IMACS and Magellan LDSS3. Each new instrument requires a special purpose code that makes crosstalk corrections (if needed) and puts the raw data and associated headers into the form expected by the data management system. In addition, each instrument requires a specific model for astrometric field distortions. Thereafter, the processing and calibration are common among all instruments.

The DESDM system will be used at NCSA for the processing, calibration and serving of DES data. The NCSA team is led by Don Petravick, who took over as the lead for the DESDM project in 2010 after Joe Mohr moved from the University of Illinois to Ludwig-Maximilians University in Munich in Fall 2009. The installations in Munich and elsewhere will be helpful as a backup, as a platform for continued development of more advanced astronomy algorithms, and as a facility to contribute during large scale reprocessing of the data. Starting in September we will support the commissioning and science verification phases of DECam on the Blanco 4m at CTIO, and thereafter we expect to move directly into the first season of science operations.

## ACKNOWLEDGEMENTS

The DESDM team acknowledges support from National Science Foundation through awards NSF AST 07-15036, NSF AST 08-13543 as well as significant seed funding provided by the National Center for Supercomputing Applications and the University of Illinois Department of Astronomy, the College of Language Arts and Science, and the Vice Chancellor for Research. DESDM activities in Munich have been supported by the Ludwig-Maximilians University and the Excellence Cluster Universe, which is supported by the Deutsche Forschungsgemeinschaft (DFG).

Funding for the DES Projects has been provided by the U.S. Department of Energy, the U.S. National Science Foundation, the Ministry of Science and Education of Spain, the Science and Technology Facilities Council of the United Kingdom, the Higher Education Funding Council for England, the National Center for Supercomputing Applications at the University of Illinois at Urbana-Champaign, the Kavli Institute of Cosmological Physics at the University of Chicago, Financiadora de Estudos e Projetos, Fundação Carlos Chagas Filho de Amparo à Pesquisa do Estado do Rio de Janeiro, Conselho Nacional de Desenvolvimento Científico e Tecnológico and the Ministério da Ciência e Tecnologia, the German Research Foundation sponsored cluster of excellence "Origin and Structure of the Universe" and the Collaborating Institutions in the Dark Energy Survey.

The Collaborating Institutions are Argonne National Laboratories, the University of California at Santa Cruz, the University of Cambridge, Centro de Investigaciones Energeticas, Medioambientales y Tecnologicas-Madrid, the University of Chicago, University College London, DES-Brazil, Fermilab, the University of Edinburgh, the University of Illinois at Urbana-Champaign, the Institut de Ciencies de l'Espai (IEEC/CSIC), the Institut de Fisica d'Altes Energies, the Lawrence Berkeley National Laboratory, the Ludwig-Maximilians Universität, the University of Michigan, the National Optical Astronomy Observatory, the University of Nottingham, the Ohio State University, the University of Pennsylvania, the University of Portsmouth, SLAC, Stanford University, and the University of Sussex.